\documentclass[aps,twocolumn,preprintnumbers,%
superscriptaddress,floatfix]{revtex4}
\usepackage[dvips]{graphics}
\usepackage{amsmath,amsfonts,bm}
\usepackage{epsfig}
\begin{document}
\newcommand{\be}{\begin{eqnarray}}
\newcommand{\ee}{\end{eqnarray}}
\def\lsim{\mathrel{\rlap{\lower3pt\hbox{\hskip1pt$\sim$}}
     \raise1pt\hbox{$<$}}} 
\def\gsim{\mathrel{\rlap{\lower3pt\hbox{\hskip1pt$\sim$}}
     \raise1pt\hbox{$>$}}} 
\def\N{${\cal N}\,\,$}
\def\prl{Phys. Rev. Lett.}
\def\np{Nucl. Phys.}
\def\pr{Phys. Rev.}
\def\pl{Phys. Lett.}
\def\la{\langle}\def\ra{\rangle}
\def\del{\partial}
\def\calL{\cal L}\def\calK{\cal K}
\def\hatn{\hat{n}}\def\Amu{{\cal A}_\mu}\def\A{{\cal A}}
\newcommand\<{\langle}
\renewcommand\>{\rangle}
\renewcommand\d{\partial}
\newcommand\LambdaQCD{\Lambda_{\textrm{QCD}}}
\newcommand\Tr{\mathrm{Tr}\,}
\newcommand\+{\dagger}
\newcommand\g{g_5}
\def\bi{\bibitem}

\newcommand{\msun}{\mbox{$M_\odot$}}
\newcommand{\rsun}{\mbox{$R_\odot$}}

\title{Kaon Condensation, Black Holes and Cosmological Natural Selection}
\author{G.E. Brown}
 \affiliation{ Department of Physics and Astronomy, SUNY, Stony Btook, NY 11794, USA}
\author{Chang-Hwan Lee}
 \affiliation{Department of Physics, Pusan National University,  Busan 609-735, Korea}
\author{Mannque Rho}
 \affiliation{ Institut de Physique Th\'eorique,  CEA Saclay, 91191 Gif-sur-Yvette C\'edex, France}

\begin{abstract}
It is argued that a well measured double neutron star binary in
which the two neutron stars are more than 4\% different from each
other in mass or a massive neutron star with mass $M\gsim 2\msun$
would put in serious doubt or simply falsify the following chain of
predictions: (1) nearly vanishing vector meson mass at chiral
restoration, (2) kaon condensation at a density $n\sim 3n_0$, (3)
the Brown-Bethe maximum neutron star mass $M_{max}\approx 1.5\msun$
and (4) Smolin's `Cosmological Natural Selection' hypothesis.
\end{abstract}

\date{\today}

\newcommand\sect[1]{\emph{#1}---}

\maketitle
\sect{I. Introduction:}
 Brown and Bethe (BB)~\cite{Bro94} proposed a simple scenario on how black holes are formed
in stellar collapse in the presence of kaon condensation in dense
compact-star matter and deduced the maximum stable neutron star mass
$M^{BB}_{max}\simeq 1.5\msun$ above which all massive stars will
collapse to black holes. This scenario which predicted about five
times more light mass black holes than the standard scenario without
kaon condensation~\cite{BBSN87,LPB07} has recently been given a
strong support by more rigorous calculations and
observations~\cite{BLR-kaon07}.

In this note we would like to point out an intricate web of relations between the critical density at which kaons can condense and the equation of state of dense compact-star matter, the maximum stable neutron star mass, `maximized' number of black holes formed in stellar collapse and  the humongous number of `pocket' universes resulting from bouncing of the singularities associated with the black holes as envisioned in `cosmological natural selection (CNS).' This is a wide-ranging set of connections involving nuclear physics, astrophysics and cosmology, the principal merit of which is that each connection is falsifiable by observations, either nuclear or astrophysical.

The predictions are these:
(a) In an alternative view to the `anthropic principle' in the  eternal inflation scenario on the landscape of the multiverse, Smolin proposes the theory of `cosmological natural selection (CNS)'~\cite{Smo97} which predicts as one of its clear-cut falsifiable predictions: ``Find a neutron star whose mass appreciably exceeds $M^{BB}_{max}$. Then it will count against the CNS scnario"\cite{Smo97,Smo04}. This prediction  follows from the requirement of the theory that the number of black holes in this universe be {\em maximized} without changing the parameters of the Standard Model.
%
(b) On the other hand, the argument for the Brown-Bethe maximum neutron star mass $M_{max}\simeq 1.5\msun$  relies on  the condensation of kaons at $\sim 3$ times the ordinary nuclear matter density $n_0$ as the {\em  first and last} phase transition in compact-star matter as the density is increased beyond $n_0$. The prediction~\cite{BLR-kaon07} is:``Find a well measured double neutron star binary in which the two neutron stars are more than 4\% different from each other (modulo some small additional shift by He red giant) in mass. Then the BB theory will be falsified." This prediction  requires that kaons condense at not too low a density $< 3n_0$ and at not too high a density $>3 n_0$.
%
(c) Kaon condensation at $\sim 3n_0$ which gives an equation of state (EOS) that when put in Tolman-Oppenheimer-Volkov equation, leads to the maximum stable neutron star mass of $\sim 1.5\msun$ has been predicted in several different ways, among which the most solid argument that we will rely on comes from low-energy effective field theory for QCD anchored on hidden local symmetry (HLS)~\cite{HY:PR}. The HLS theory predicts that at some high density in the vicinity of chiral restoration,  the hidden gauge coupling goes to zero due to what is known as `vector manifestation (VM) of chiral symmetry' and hence so do hadronic interactions. Kaons condense in this `weakly interacting' region. The falsifiable prediction is this: ``Find a neutron star of mass $\gsim 2 \msun$, whether in binary or otherwise, then it falsifies the VM of HLS theory, which in turn falsifies the kaon condensation at $\sim 3n_0$." This prediction results because without the VM, there can be strong repulsion at a density $> n_0$ built through many-body forces which will push kaon condensation to much higher density and could allow transition to color-superconducting  quark star.

\sect{II. Cosmological Natural Selection:}
As an explanation for the complexity of the multiverse, Smolin~\cite{Smo97} puts forward cosmological natural selection mechanism as to how we live in the particular unverse we are in.
The idea of the CNS scenario is that in addressing the multiverse structure in cosmology which appears to be also supported by string/M theory, the mechanism for universe reproduction  is a bouncing black hole singularity that leads to a new expanding region of space time, i.e. a universe, behind the horizon of every black hole. Here while accepting a rich landscape of the kind string theory suggests, the CNS scenario differs from  the anthropic principle with eternal inflation -- which states that the universe is eternally inflating, endlessly spawning `pocket universes' -- in that it relies on  the maximized formation of black holes. Among a variety of conditions for forming many black holes, there is one which concerns us, that is that, the upper mass limit of neutron stars be as {\em low} as possible~\cite{Smo04}

We are not in a position, nor is it our objective, to address the basic questions associated with the scenario, such as how black hole singularities bounce and reproduce stupendous numbers of new universes, how to resolve the information loss in the reproduction, how the size of a black hole relates to the number or size of universes, whether the CNS is an alternative or superior to the anthropic principle with eternal inflation etc. The dispute with pros and cons for both will wage on~\cite{debate}, on which we have nothing to say. What we can address is one of the falsifiable predictions for the CNS, i.e. the upper mass limit of neutron stars that maximizes the number of black holes formed.

\sect{III. Maximization of Black Holes $\approx$ Maximization of the Entropy:}
When the Fe core of a large star drops into a neutron star, the entropy per nucleon in the Fe core is about unity~\cite{Bet79}. A solar mass of such matter has $\sim 10^{57}$ nucleons, so this means an entropy of $\sim 10^{57} k_B$.  Bekenstein \cite{Bec73} showed that black holes had entropies of
\be
S_H = \frac{k_B}{4} \times \frac{\rm surface\; area\; of\; horizon}{\rm (Planck\; length)^2}
\ee
which for a Schwarzschild black hole means
\be
S_H= 1.05 \times 10^{77} k_B \left(\frac{M}{\msun}\right)^{2}.
\ee
Thus, when Fe cores go into black holes, the entropy is increased by a factor of $10^{20}$ per particle (although the particles have gone into {\it `nothingness'} when they went into a black hole. Still, the entropy is on the surface (event horizon)).

Now a fundamental law of nature is that a system moves toward equilibrium in such a way as to maximize the entropy. Therefore,  the maximum number of black holes does the best, so far, in moving towards equilibrium.

In maximizing the number of black hole formations, there are two questions to be posed: One,  what is the lowest upper mass of neutron star?; two,  what is the maximum upper mass ?

\sect{IV. Carbon Abundance and the Lowest Upper Mass (LUM)}
We start with the LUM. Can it be lower than say 1.4 $\msun$?

First of all,  we cannot lower the maximum neutron star mass much because the existing Hulse-Taylor pulsar has mass $1.44\msun$. But if we could, would we want to? The answer is no, because lowering it would get in the way of carbon (and, therefore, oxygen) production. In addition to minimizing the upper limit of neutron stars, at the same time the production of massive stars must be extremized, in order to produce the maximum number of black holes. The production of massive stars depends on processes of cooling that involve carbon and oxygen.
%

The argument on the carbon abundance goes like this~\cite{Bro01}: Carbon is formed out of 3 $\alpha$-particles, the process being enhanced when two of the $\alpha$-particles go into an unbound resonance stage of ${^8}{\rm Be}$,
\be
\alpha + \alpha + \alpha\longrightarrow \alpha + {^*}{^8}{\rm Be} \longrightarrow {^{12}}{\rm C}.
\label{eq-a}\ee
Carbon is removed by a two-body process
\be
{^{12}}{\rm C} + \alpha \longrightarrow {^{16}}{\rm O}.
\label{eq-b}\ee
Now any carbon not removed by this process must burn in processes such as
\be
{^{12}}{\rm C} + {^{12}}{\rm C} \longrightarrow {^{24}}{\rm Mg},
\label{eq-c}\ee
the ${^{24}}{\rm Mg}$ going into a variety of elements. The point that is important for us is that whereas the reactions (\ref{eq-a}) and (\ref{eq-b}) proceed at a temperature of 20 keV, reaction (\ref{eq-c}) requires 80 keV because of the much greater Coulomb barriers. Now this latter process (\ref{eq-c}) at such an energy is accompanied by loss of entropy, because neutrino pairs $\nu +\bar\nu$ have an emission cross section which goes as $T^{11}$, and the temperature is high enough so that the amount of entropy carried off is $\sim 1/3$ of that in the original star.
However, the basically three-body process (\ref{eq-a}) goes as $\rho^2$ and the removal of ${^{12}}{\rm C}$ in (\ref{eq-b}) goes as $\rho$ where $\rho$ is the density. As the zero age main sequence (ZAMS) mass in stars increases, the density goes steadily downward, favoring (\ref{eq-b}) against (\ref{eq-a}). In our Galaxy, the two processes become equal at ZAMS mass $\sim 18\msun$ (just the mass of Sanduleak-69$^\circ$202, the progenitor of SN1987A). The ${^{12}}{\rm C}$ has been removed nearly as quickly as it has formed.

Now the point is that black holes  start forming at a ZAMS mass just slightly above the minimum in carbon abundance. When the central carbon abundance drops below 15\%, there is not enough energy for convective (steady) carbon burning. Thus, entropy is no longer carried off by reaction (\ref{eq-c}). But the total entropy increases with increasing ZAMS mass. This can only happen through increasing the mass of the Fe core, which has an entropy of $\sim k_B$ per nucleon \cite{Bet79}. This increase in the mass of the Fe core causes the star to go into a black hole.

In fact, from the mass cut necessary to produce the $\sim 0.075\msun$ of Ni in SN1987A explosion, Bethe and Brown~\cite{BBSN87} reconstructed the explosion, determining the upper limit on the gravitational mass of the progenitor as 1.56 $\msun$. Reconstruction of how this progenitor went into a black hole was made by Brown and Weingartner~\cite{brown-wein}.

Were the lowest upper neutron star mass as low as, say, $1.4\msun$,  we would have more black holes but we know that the Hulse-Taylor binary with pulsar of $1.44\msun$ exists, so Nature would rule it out. Furthermore even if it were possible, the black holes will begin forming at a ZAMS mass less than $18\msun$. This would begin cutting out the ${^{12}}{\rm C}$ that exists up to the ZAMS mass at which black holes are born.
%

There is, however, a complication. Namely, stars of ZAMS mass $18-30\msun$ do not fall immediately into black holes as they collapse. SN1987A lived long enough to emit neutrinos for 12 seconds. Woosley arrived at his value of 170 keV barns for the $^{12}$C($\alpha,\gamma$)$^{16}$O process by reproducing observed abundances of ZAMS $25\msun$ to determine this value. His value is close to the Kunz et al. \cite{Kun01} $165\pm 50$ keV barns, the best measured value. The point is that ZAMS mass stars of up to $30 \msun$ have time to burn and expel metals in the ``delayed explosion" before they fall into black holes. This is discussed in detail and in length by Brown \& Bethe \cite{Bro94}.

\begin{figure}[hb]
\begin{center}
{\epsfig{file=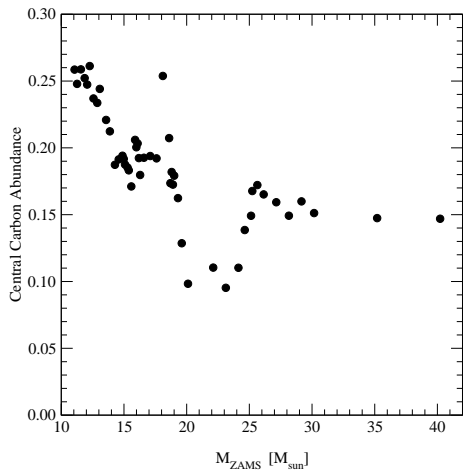,height=3.0in}}
\vskip -0.5cm
\caption{ Central carbon abundance at the end of He core burning for
``clothed" (single) stars as function of ZAMS mass \cite{Bro01}.
The rapid drop in the central carbon abundance at ZAMS mass $M_{\rm ZAMS}\sim 18\msun$ signals the disappearance of convective carbon burning.  }
\label{fig1}
\end{center}
\end{figure}

Schaller et al. \cite{Sch92} used the old Caughlan-Fowler \cite{Cau88} value of 100 keV barns for the $^{12}$C($\alpha,\gamma$)$^{16}$O reaction and obtained the sudden drop in the central carbon abundance at ZAMS $25\msun$.
%
%
But the Caughlan-Fowler \cite{Cau88} value of the $^{12}$C($\alpha,\gamma$)$^{16}$O reaction is now known to be too small and with the correct higher value the central carbon abundance curve is as in Fig.~\ref{fig1}.

The minimum ZAMS mass for black hole production below $18\msun$  would mean that the minimum in $^{12}$C production would come in the region of the much more copious low-mass stars. (The Salpeter mass population function goes as $dN/dM \propto M^{-2.35}$.)
This would mean much less $^{12}$C production in the Galaxy.

\sect{V. The BB Scenario and Kaon Condensation:}
Now what about the maximum upper mass? This is where the phase structure of compact-star matter figures. The BB scenario is based on kaon condensation as the most crucial phase transition in compact stars. Positing that it takes place at $n\sim 3n_0$, BB obtain the maximum upper mass to be $M_{max}\approx 1.5\msun$.

There have been a flurry of theoretical activities on the phase
structure of nuclear matter ranging all the way to infinite density,
obtaining a landscape as rich as that of the water. While QCD is
predictive at asymptotic density, it is moot at the densities
relevant to compact stars. The predictions made so far in the
relevant density have been based on models, the validity of which
has been confirmed neither theoretically nor empirically. The only
phase transition that can be treated with some confidence is
fortunately the first phase transition that takes place as the
density increases beyond the nuclear matter density, and that is
kaon condensation. What makes kaon condensation accessible is that
it can be treated by effective field theories anchored on chiral
symmetry. It is our key thesis that this transition determines the
fate of compact stars either as neutron stars or black
holes~\cite{BLR-kaon07}. If kaons do condense at $\sim 3n_0$ as
predicted, then quark matter -- with or without color
superconductivity -- that can appear at higher densities in a star
with $M\gsim M_{max}^{BB}$ would be buried in the `nothingness' of
black holes.

There are three density regimes where effective field theories can
be exploited for quantitative calculations: (1) in the vicinity of
$n=0$ for which chiral perturbation theory applies, (2) near $n=n_0$
for which the Landau Fermi-liquid fixed point theory is well
established and (3) near $n\approx 4n_0$  at which chiral symmetry
is restored and HLS has its vector manifestation fixed point. What
is significant is that kaon condensation takes place close to the
vector manifestation fixed point, so it is most astute to approach
it by hidden local symmetry theory. It also lends itself to a
clear-cut verification or  refutation. What hidden local symmetry
theory predicts is that near the critical density $n_c$, all the
masses of the hidden gauge particles, the principal degrees of
freedom in the energy regime concerned,  become tiny -- zero in the
chiral limit--  as a consequence of the hidden gauge coupling
constant going to zero.

Fluctuating from the VM fixed point,  kaon condensation turns out to
be quite simple~\cite{BLPR-kaoncon,BLR-kaon07}. Imagine embedding  a
$K^{-}$ in a neutron-star matter. As the density increases, the kaon
will feel the attractive force from the medium and its mass will
drop. In HLS theory, the kaon mass will go to zero at about the same
point where the vector meson mass goes to zero. There are no kaons
in neutron-star matter but there are electrons, whose chemical
potential will increase as the matter density increases. When the
electron chemical potential increases to the level~\cite{footnote1}
at which the mass of a kaon in such dense matter would have dropped
to the same level, the electron decay to a kaon  by the weak
interaction, $e^-\rightarrow K^- +\nu_e$, becomes energetically
favorable and will take place because kaons are bosons and so tend
to condense, preferably in the zero-momentum state, i.e. s-wave. At
this point, the flow of the matter toward the VM fixed point will be
stopped and will be taken over by a new form of matter, making the
EOS of the matter softer. This happens at near $n\approx
3n_0$~\cite{BLR-kaon07} giving the BB maximum mass
$M^{BB}_{max}\approx 1.5\msun$. Were it much lower, then both the
Hulse-Taylor pulsar and the carbon abundance  mentioned above would
falsify the scenario.

But could it be $n > 3n_0$ or $M_{max} > 1.5\msun$?

Although most well-measured binary pulsars satisfy the bound of $M_{max}^{BB}=1.5\msun$, there are reported cases of compact-star masses that exceed the BB maximum mass. Up until recently, the most serious case against the BB scenario was PSRJ0751$+$1807, a neutron star in a binary with white dwarf,  with mass $2.1^{+0.20}_{-0.20} \msun$
\cite{Nice05} which had spurred a large number of works purporting to rule out the  kaon condensation at as low a density as $n\sim 3n_0$ as well as to provide support for quark stars with or without color superconductivity.  This would have been a clean falsification of the BB theory as well as the CNS idea. However a recent analysis by the same group lowered the mass to $1.26^{+0.14}_{-0.12} \msun$ (see D. Nice, talk in {\it 40 Years of Pulsars}, Aug. 12-17, McGill University, http://www.ns2007.org). There are other cases of higher mass neutron stars but there are reasons to believe that as they stand, they cannot be taken as a serious negative evidence. This matter is discussed in depth in \cite{BLR-kaon07}. At present, it seems fair to conclude that  there is no ``smoking-gun" evidence against the BB scenario.

A firm observation of any type of a neutron star whose mass is
greater than $M_{max}^{BB}$ or to be safe $\gsim 2\msun$ would
present a serious obstacle to the BB and CNS scenarios. An
interesting case where the web of connections can be tested is
possible gravitational wave signal from Laser Interferometer Gravitational-Wave Observatory (LIGO). Suppose kaon
condensation fails to send a massive star into a black
hole~\cite{footnote2}. Then a three-flavor crystalline color
superconducting (CCS) phase of QCD can develop as the ground state
of cold matter inside compact-star matter. Such a phase can have
solid state properties, e.g., lattice structure and shear modulus,
and multipolar deformations in gravitational
equilibrium~\cite{ippolitoetal}.  The breaking strain of solid in
the interior of a spinning neutron star with shear moduli up to 1000
times that of conventional neutron stars~\cite{mannarellietal07}
could give rise to observable gravitational wave signals. Such a
signal has been searched for in gravitational wave search from  LIGO,
but so far with no success~\cite{haskelletal}.

But suppose that there is a confirmed signal for a solid structure in the core of the star of the type predicted by the CCS . What this could mean is as follows. Model calculations show that in order for nuclear matter to make a transition to the crystalline structure,  the EOS of the nuclear matter has to be {\em maximally} stiff~\cite{ippolitoetal}. While the models used are not quantitatively trustworthy, the qualitative feature that requires the stiffness of EOS is likely robust. The stiffness required for the transition then implies that kaon condensation cannot occur up to  a density  $\gsim 7n_0$~\cite{pandha}. This would  falsify the BB scenario and put in doubt the CNS theory.  It would also falsify the vector manifestation of hidden local symmetry which predicts that the strong repulsion appearing at $n > n_0$ required for the transition from nuclear matter to the CCS state will be suppressed as the density increases. This has been confirmed in the soliton description of dense hadronic matter which is valid in the large number- of- colors ($N_c$) limit~\cite{PRV}.
Thus, we uncover a web of falsifiable connections, the vector manifestation with a vanishing vector meson mass, soft nuclear matter EOS, kaon condensation, the BB scenario for $M_{max}^{BB}$ and cosmological natural selection for maximizing black holes.

\sect{Acknowledgments} GEB was supported in part by the US
Department of Energy under Grant No. DE-FG02-88ER40388.
CHL was supported by the BAERI Nuclear R\&D program (M20808740002) of MEST/KOSEF.


\begin{thebibliography}{99}
\bibitem{footnote1}
A possible obstacle to this scenario is the appearance of negatively
charged hyperons, e.g., $\Sigma^-$ in the matter at a lower density.
How this obstacle can be avoided in delayed collapse as well as
loop-holes in the stadard arguments leading to this obstacle are
discussed in \cite{BLR-kaon07}.
\bibitem{footnote2} A star with condensed kaons might exhibit
anomalous cooling as discussed in the literature. However this would
require calculating the equation of state nonperturbatively in
hidden local symmetry theory.
\end{thebibliography}
\end{document}